\documentclass[]{interact}
\usepackage[caption=false]{subfig}
\usepackage{epstopdf}
\usepackage[natbibapa,nodoi]{apacite}
\usepackage[bookmarksnumbered,colorlinks,bookmarks,citecolor=blue,linkcolor=blue,urlcolor=blue,breaklinks,linktocpage]{hyperref}
\usepackage{xcolor}
\usepackage{listings}
\usepackage{amsmath} 
\usepackage{graphicx}
\usepackage{url}

\usepackage{amssymb}
\usepackage{epstopdf}
\begin{document}

\markboth{Lazzarini and Timoney}{}

\title{Higher-Order Frequency Modulation Synthesis}

\author{
\name{Victor Lazzarini\thanks{email: Victor.Lazzarini@mu.ie}  
and  Joseph Timoney\thanks{email: Joseph.Timoney@mu.ie}}
\affil{Maynooth University, \\ Maynooth, Ireland}
}

\maketitle
\begin{abstract}
Frequency modulation (FM) and phase modulation (PM) are well-known synthesis methods, which have been deployed
widely in musical instruments. More recently, some synthesisers have implemented direct forms of FM (as opposed to PM), allowing, at least as part of their design, for higher-order modulation topologies. However, such implementations are affected by well-known difficulties that arise in the modulation of frequency, which are normally solved by the use of PM. In this article, we analyse these problems and using a direct comparison with PM, we put forward a solution for the direct application of FM in higher-order modulation arrangements. We begin by reviewing the theory of first-order FM, contrasting it to PM. We then proceed to develop a formulation of second-order FM which is equivalent to the issue-free PM synthesis, and present a closed-form expression for the evaluation of the second-order FM spectrum. We then extend the principle to higher-order topologies,
by advancing the concept of an FM operator, analogous to the one used in PM instrument designs. From this we demonstrate that feedback FM is also a practical possibility. Finally, we complement the paper by giving a reference implementation in C++. 
\end{abstract}

\begin{keywords}
sound synthesis; non-linear distortion; frequency modulation; phase modulation; feedback;
\end{keywords}

\section{Introduction}

Linear frequency modulation (FM) as a sound synthesis technique has had a long history of development, first explored by James Tenney, followed by
Jean-Claude Risset and John Chowning~\citep{Lazzarini2023}. It was given a theoretical treatment by \cite{ChowningFM}, where it was
demonstrated to provide an economical method of producing dynamic spectra with both harmonic and inharmonic partials. It was shown to generate sounds previously only possible with the more expensive means of additive synthesis. In the equivalent, but more flexible form of phase modulation (PM), which was the actual object of analysis in Chowning's paper, the method was implemented in a very successful range of digital synthesisers, first by Yamaha~\citep{FMTheory}, then by other manufacturers. In this form it was expanded to support higher-order (or stacked), as well as feedback, modulation. Both extensions can be characterised as forms of complex (as in \emph{multi-component}) PM. 

The mathematical description of FM and PM actually stems from the early studies in radio frequency broadcasting~\citep{Bloch, Corrington}. The concept of an instantaneous frequency, as the time derivative of the phase angle, was first introduced in order to support the development of FM theory~\citep{Carlson}. In these early papers, the exact distinction between the two forms of modulation was not a concern for the authors, as the principles being developed could be implemented with either one of the methods. However, it is important to note that most of the mathematics underpinning it, arising from the theory of Bessel functions, applies directly to PM and only in a second instance to FM, as we will show in this paper. 

The subject of PM has been studied extensively since Chowning's original paper, in many cases under the misleading name of FM. A review article by Moorer showed that phase modulation is in fact a particular instance of a more wide class of nonlinear techniques, which may described by closed-form summation formulae~\citep{MoorerIEEE}. Such methods also include waveshaping~\citep{LeBrunWaveshaping}, phase distortion~\citep{LazzariniPD}, and different forms of formant synthesis~\citep{Lazzarini2017}. More recently, we have had the development of adaptive techniques, such as adaptive FM~\citep{LazzariniADFM}, which allowed phase modulation of arbitrary sources. The introduction of modified FM synthesis is also worthy of note, producing yet another variant of PM based on purely imaginary modulation indices~\citep{Lazzarini2010theory}. Recently, we have seen the introduction of the concept of Loopback FM~\citep{Smyth, Loopback}, and the question of taming exponential FM~\citep{HutchinsXFM} in analogue and digital synthesis applications has also been the object of further studies~\citep{TimoneyEFM,Nielsen}. A complete account of the state of the art of non-linear distortion synthesis techniques is found in~\cite{Lazzarini2021}.

In this paper, we first clarify the differences between FM and PM, making sure that the definitions of the two techniques are well established. Then we will proceed to discuss the question of higher-order modulation, which is realised by the use of frequency-modulated signals in a stack of modulators. This is a technique that is well understood as far as PM is concerned, but has not been fully described in spectral terms before, and in particular has not yet received a treatment in the case of FM. We begin by focusing on the specific case of second-order FM, for which an equivalent PM expression is derived. From this, we have both an implementation recipe in the form of a synthesis flowchart, and a means of deriving the resulting spectrum. Higher-order FM is then shown to be a generalisation of this particular case, with a practical implementation
through the concept of FM operators (analogous to the well-known PM operators described by \cite{FMTheory}).
To complement, we put forward a reference implementation in C++ to illustrate the principles presented earlier on.

\section{Frequency Modulation}
FM synthesis is fairly straightforward to implement. In its most general form, a signal is used to control the frequency of an oscillator, producing an output with many partials. The frequency control may be exponential or linear. In this work, we concentrate on the latter form of FM. The technique has been described in terms of a \emph{fast} vibrato, which employs modulation frequencies within the audio range ($> 20$ Hz). We can take this intuitive description as our starting point. 

Vibrato has two fundamental parameters: rate and width. The latter is determined by the amplitude of the modulating signal and the former by its frequency.  We can also define the width as the maximum absolute deviation from a centre frequency. While at sub-audio rates the result of vibrato is a certain fluctuation of pitch, as the modulation frequency and width increases, the carrier output signal ceases to be perceived as a pure sinusoid and becomes a waveform whose spectrum features a number of partials. It is necessary that enough modulation is applied, making the oscillator instantaneous frequency negative at times.

We can represent a sinusoidal FM carrier signal $c(t)$ using the following expression,

\begin{equation}\label{eq:fm}
c(t) = \cos\left(2\pi \int_{0}^{t} f_c + m(x) dx \right).
\end{equation}

To facilitate the discussion, we can set the modulator to $m(t) = d \cos(2\pi f_m t)$, a sinusoid with amplitude
$d$ and frequency $f_m$. We then have a modulation frequency $f_m$ and a carrier frequency deviation $d$,
along with the carrier frequency $f_c$, as the main parameters of FM, 

\section{Phase Modulation}

To formulate an equivalent form of PM, 
we can first rewrite Eq.~\ref{eq:fm} as 

\begin{equation}\label{eq:pm}
c(t) = \cos(\phi(t)),
\end{equation}

\noindent that is, using a time-varying phase signal $\phi(t)$ as a driver to the carrier signal.
We can put this function instead in terms of sinusoidal phase modulation (PM),

\begin{equation}\label{eq:pmod}
\phi(t) =  2\pi f_c t  + z \sin(2\pi f_m t)
\end{equation}

The advantage of the PM representation is twofold. First, we 
have a measure of the amount of modulation, $z$, that, as we can demonstrate, does not depend on the
modulation frequency; and, second, we can take advantage of the Jacobi-Anger 
expansion~\citep{Watson}, to determine the spectrum of the phase modulation signal,

\begin{equation}\label{eq:jacobi-anger}
\begin{split}
e^{\pm jz\sin(\theta)}  &= J_{0}(z) + 2 \sum_{n=1}^{\infty} J_{2n}(z)\cos\left(2n\theta\right) \\
& \pm 2j \sum_{n=0}^{\infty} J_{2n+1}(z)\sin\left([2n+1]\theta\right),
\end{split} 
\end{equation}

\noindent where $J_n(z)$ is the Bessel coefficient of order $n$. From this equation, we can observe
that $z$ is directly involved in determining the amount of energy spread from
the carrier partial to the various sidebands.

\subsection{Equivalence to FM}\label{sec:fmvpm}

Now in order to connect this to the FM expression of Eq.~\ref{eq:fm}, we can find 
the corresponding instantaneous carrier frequency as the derivative of the phase
signal $\phi(t)$ in Eq.~\ref{eq:pmod}~\citep{Moore:1990},

\begin{equation}
\dot{\phi}(t) = 2\pi \left [f_c  + z f_m \cos(2\pi f_m t) \right]. 
\end{equation}

The quantity $z f_m$, the product of the modulation frequency and the phase modulation amount
is equivalent to the frequency deviation $d$ employed in the FM signal. We use the term 
\emph{modulation index} to characterise $z$,

\begin{equation}\label{eq:deviation_fm}
z =  \frac d {f_m}
\end{equation}

As shown before, the modulation index determines the bandwidth, or spread of the resulting FM/PM spectrum.
From Eq.~\ref{eq:jacobi-anger}, we can derive

\begin{equation}\label{eq:pm_synthesis}
\cos\left(\omega + z\sin(\theta)\right) = \sum_{n=-\infty}^{\infty} J_n(z)\cos\left(\omega + n\theta\right).
\end{equation}

\begin{figure}[htp]
\begin{center}
\includegraphics[width=.5\columnwidth]{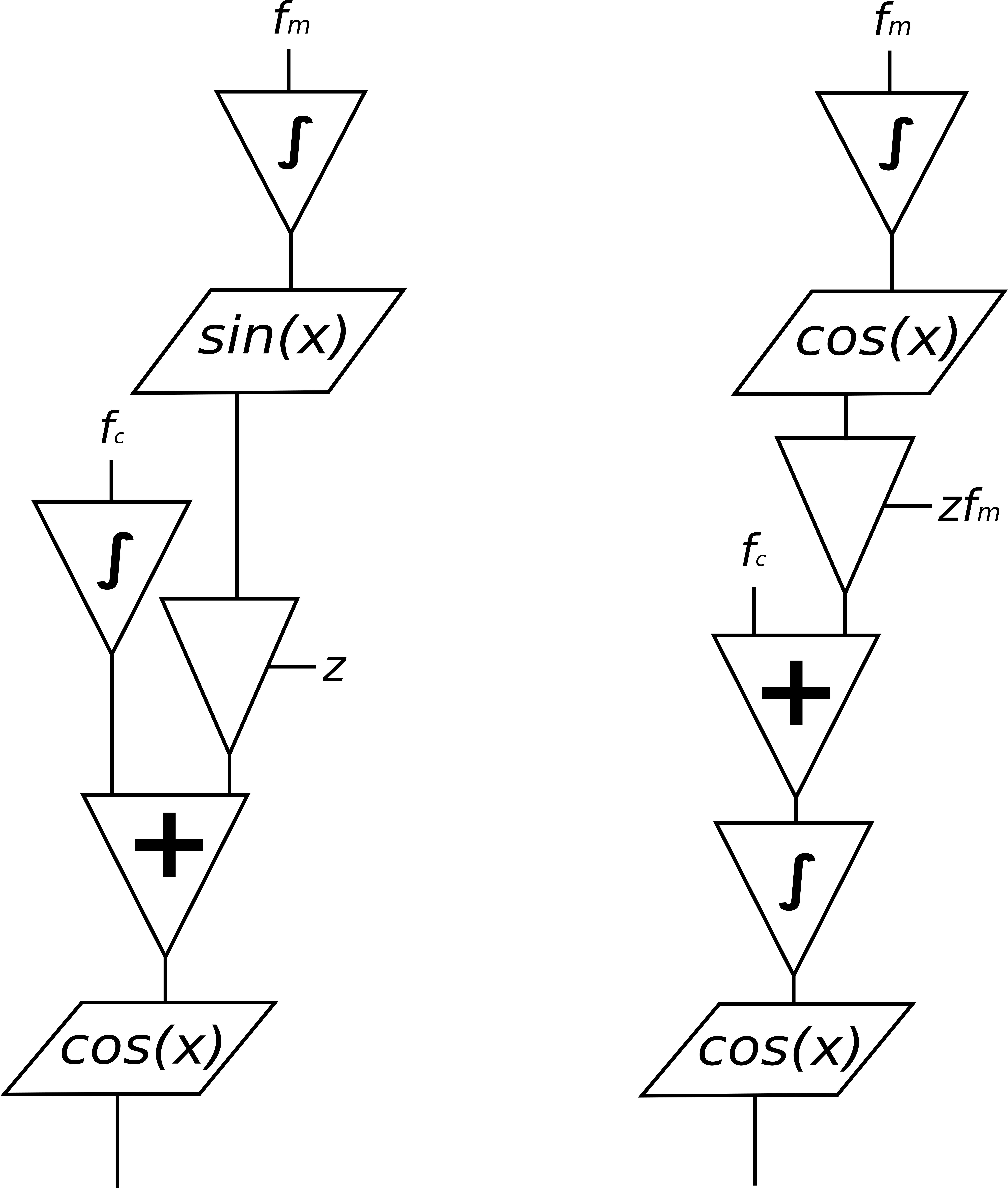}
\caption{Flowcharts for PM (left) and FM (right).}
\label{fig:fmpm}
\end{center}
\end{figure}

\noindent This demonstrates that the amplitude of the frequency modulator, $d$, cannot alone define
a measure of the amount of modulation applied to the carrier signal. On the other hand,
as we noted, the amplitude of the phase modulator can be used directly to determine the output spectrum.
Setting $\omega = 2\pi f_c t$ and  $\theta = 2\pi f_m t$, we can see that the spectrum of
Eq.~\ref{eq:fm} is composed of partials at $f_c \pm n f_m$ Hz, which are scaled by the
corresponding Bessel function coefficient $J_n(z)$. We now have a mechanism to 
represent the FM in terms of the PM, which provides a clear route for analysis.
A comparison between the FM and PM flowcharts is shown in Fig.~\ref{fig:fmpm}.

\section{A Theory of Second-Order FM}

 The case of Eq.~\ref{eq:fm} is that of first-order modulation,
consisting of one modulator and one carrier oscillator. We now consider the arrangement whereby this is increased
to a higher order, that is, where two or more modulation stages are present. An FM carrier wave is then 
used as a modulator to a subsequent carrier, and this may be used again as a modulator, and so on. 
The output signal is the result of several orders of modulation. It has been claimed in the literature
that, unlike PM, FM synthesis cannot be implemented in higher-order topologies~\citep{PinkstonFM}, but
as we will demonstrate, that is the not the case. Most of the difficulties arise from a simplistic approach
to the implementation of FM synthesis that does not take into account the differences we have discussed
in section \ref{sec:fmvpm}. It is the case that some synthesisers implement direct forms of FM allowing the possibility of higher-order modulation topologies (and even feedback) (see for instance \cite{Summit}). However, there are deficiencies in these designs, which we address in this paper.

We first need to consider the amount of modulation required at each stage. A na\"{i}ve approach,
following directly from the single-level example, would lead us to apply simply the product of an index of
modulation $z_n$ and modulation frequency $f_{m_n}$ to determine the frequency deviation at each stage $n+1$. 
Not only this is incorrect from a mathematical point of view, but also, depending on the ratio of the modulation frequencies involved, 
we will observe some problematic results. Consider the case of a second-order modulation arrangement, 
where we have a first order modulator with frequency $f_{m_0}$, modulating a second-order modulator
with frequency $f_{m_1}$, which then modulates a carrier oscillator with frequency $f_{c}$,
where we set $f_{m_0} = f_{m_1} = f_c$. We know from Eq.~\ref{eq:pm_synthesis} that the resulting spectrum at
the first order stage may contain a term at $0$ Hz (DC), which is scaled by $J_{-1}(z)$. When this modulation 
signal is then applied to the frequency $f_c$ of a carrier oscillator at the next stage, the DC term is added to $f_c$. This results in a partial whose frequency drifts in accordance to changes in the amplitude
of the DC term of the modulation signal. Such an effect becomes hard to control and is undesirable in most practical applications. It is of  course possible to select modulation 
frequency ratios that do not result in DC terms, and also to produce spectra with a strict
$-\pi/2$ phase at 0 Hz, however these solutions are not general enough to support a
theory of higher-order FM synthesis. 

Note that such issues do not occur in PM, since any DC offset is translated as a phase shift,
rather than a carrier frequency drift. For this reason, it is generally much more flexible to
adopt PM as a general method for higher-order modulation. However, it is still possible to
develop a solution based on the principles outlined earlier. For this, we need to consider that 
the integration involved in FM synthesis (cf Eq.~\ref{eq:fm}) requires that a periodic time-varying 
deviation is applied to the signal. Since the instantaneous frequency of an FM signal is 
time-varying, it implies the presence of an amplitude modulation term following 
integration. 

For example, let's consider a modulator signal $m_1(t)$ whose frequency $f_{m_1}$ is itself modulated by
a sinusoidal signal $m_0(t)$, whose frequency is $f_{m_0}$. If we want to apply an index of
modulation $z_0$, then according to Eq.~\ref{eq:deviation_fm}, we need to set 
the $m_0(t)$ signal amplitude $d_0$ to $z_0 f_{m_0}$. The time-varying frequency $f(t)$ of
the modulator $m_1(t)$ is then 

\begin{equation}
f(t)  = f_{m_1} + z_{0} f_{m_0} m(t),
 \end{equation}

Therefore, also according to eq.~\ref{eq:deviation_fm}, 
we need to employ the following time-varying deviation, 
with an appropriate value of $z_1$,

\begin{equation} \label{eq:deviation_stack}
d_1(t) = z_1 f(t) = z_1 [f_{m_1} + z_{0} f_{m_0} m(t)],
\end{equation}

\noindent in order to produce the correct frequency modulation signal. 

\subsection{Second-order FM Analysis}\label{sec:theory}

We can now apply these principles to the typical case of a second-order FM synthesis arrangement using two
sine wave oscillators. The solution developed for this case can then be expanded to provide methods
for higher-order modulation topologies. Using the notions developed above, we can now describe 
the correct form of second-order FM as

\begin{equation} \label{eq:fmstacked}
\begin{split}
&m_0(t) =  \cos(2\pi f_{m_0} t)\\
&m_1(t) = \cos\left(  2\pi\int_{0}^{t} f_{m_1} + z_0  f_{m_0} m_0(x)dx\right) \\
&c(t) =  \cos\left(2\pi \int_{0}^{t} f_c + z_1 [f_{m_1} + z_0 f_{m_0} m_0(x)] m_1(x) dx \right). \\
\end{split}
\end{equation}

We can now demonstrate how these equations are equivalent to the typical form
of second-order PM. To do this, we begin by reworking the first-order modulation 
as a PM expression,

\begin{equation}\label{eq:fm+pm}
\begin{split}
&m_0(t) = \sin(2\pi f_{m_0} t)\\
&\phi(t) = \cos(2\pi f_{m_1} t + z_0 m_0(t)) \\
&c(t) =  \cos\left(2\pi \int_{0}^{t} f_c + z_1 [f_{m_1} + z_0 f_{m_0}\cos(2\pi f_{m_0} x)] \phi(x) dx\right). \\
\end{split}
\end{equation}

The next step is to replace $\phi(.)$  in the carrier signal 
equation,

\begin{equation}
\begin{split}
c(t) = \cos\Bigg(2\pi \int_{0}^{t} &f_c  + z_1 [f_{m_1} + z_0 f_{m_0} \cos(2\pi f_{m_0}x)] \times\Bigg.\\
&\Bigg. \cos\left(2\pi f_{m_1} x + z_0\sin(2\pi f_{m_0}x)\right) dx\Bigg),
\end{split}
\end{equation}

\noindent which translates as the following expression describing second-order PM

\begin{equation}\label{eq:pm_stack}
c(t) = \cos\left(2\pi f_c t + z_1\sin\left(2\pi f_{m_1} t + z_0\sin(2\pi f_{m_0} t)\right) \right).
\end{equation}

\begin{figure}[htp]
\begin{center}
\includegraphics[width=.5\columnwidth]{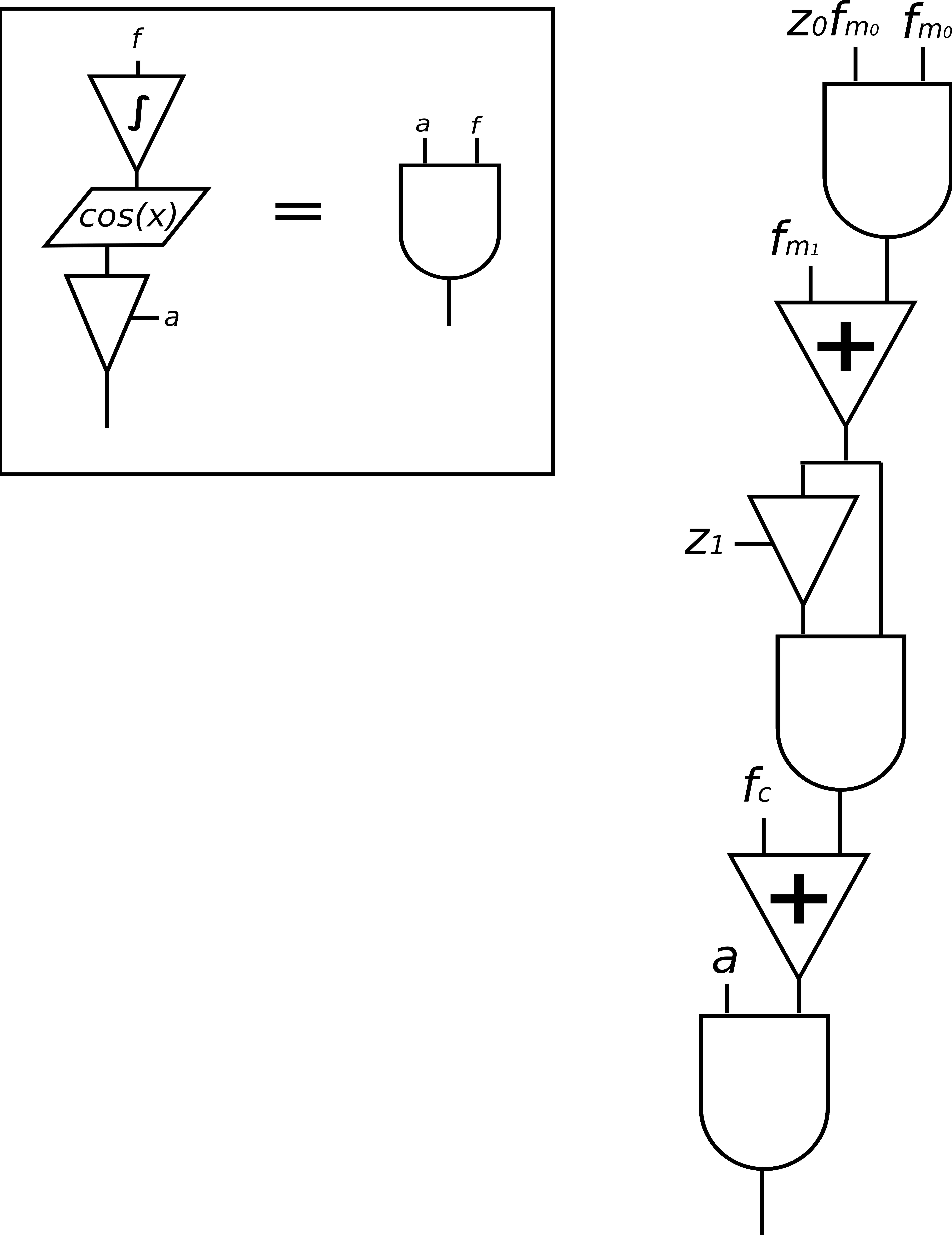}
\caption{Second-order FM flowchart.}
\label{fig:stacked}
\end{center}
\end{figure}

The equivalent FM topology (Eq.~\ref{eq:fmstacked}) can be implemented 
using the flowchart shown in Fig.~\ref{fig:stacked}. In this, we see that in order to 
implement stacked FM we need to take account of the amplitude modulation effects that arise
from employing a modulated input, as per Eq.~\ref{eq:deviation_stack}.

\begin{figure}[htp]
\begin{center}
\includegraphics[width=.75\columnwidth]{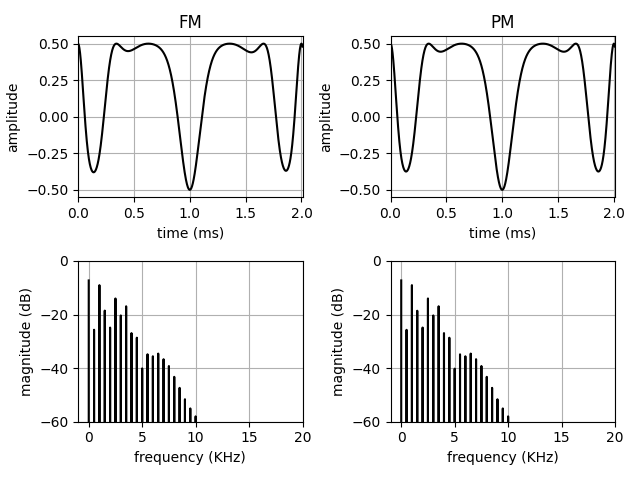}
\caption{Second-order FM (left) and PM (right) waveforms and spectra from Eq.~\ref{eq:fmstacked} (Fig~\ref{fig:stacked}) and Eq.~\ref{eq:pm_stack}, respectively, with $f_c = f_{m_0} = f_{m_1} = 500$ Hz, $z_0=3$, and $z_1=2$}
\label{fig:fm-pm}
\end{center}
\end{figure}

Waveforms and spectra examples produced by second-order FM and PM are shown
to be equivalent in Fig.~\ref{fig:fm-pm}. These signals were produced using the approach
described by Eq.~\ref{eq:fmstacked} and Fig~\ref{fig:stacked}, in the case of FM, and the
corresponding PM expression given in Eq.~\ref{eq:pm_stack}. This demonstrates that it is indeed possible to
use a second-order FM topology to produce a spectrum that is similar to second-order PM.  

\subsection{Spectrum}

With its equivalent PM form, we can now derive an expression for the second-order FM spectrum. Using
Eq.~\ref{eq:jacobi-anger} we rewrite Eq.~\ref{eq:pm_stack} as

\begin{equation}
c(t) =  \cos\left( 2\pi f_c t + z_1 \sum_{n=-\infty}^{\infty} J_n(z_0)\sin\left(2\pi [f_{m_0}  + n f_{m_1}] t\right)\right).
\end{equation}

From this equation we can now use a derivation of the spectrum of complex PM~\citep{LeBrunFM}. In order to make this 
more meaningful, we assume that the modulation signal spectrum contains only $K$ sidebands 
with significant energy (rather than a theoretically non-bandlimited spectrum). The spectrum of
second-order FM synthesis is then defined as

\begin{equation}\label{spectrumeq}
\begin{split}
\sum_{n_{-K}=-\infty}^{\infty} \ldots \sum_{n_{K}=-\infty}^{\infty}
&\left[\prod_{k=-K}^{K} J_{n_k}\left(z_1 J_k(z_0)\right) \right]\\
&\cos\left(2 \pi [f_c  + \sum_{k=-K}^{K} k(f_{m_1} + n_k f_{m_0})] t \right).
\end{split}
\end{equation}

The value of $K$ is dependent on the modulation indices employed and will increase as more
modulation is inserted into the signal. High values of $K$ may produce audible aliasing in 
digital implementations, depending also on the modulation and carrier frequencies as well as
the sampling rate.

\section{Higher-order Modulation}
The method developed here for second-order modulation may be expanded to higher-order arrangements,
as needed. The modulation signal at each level is inserted into both the amplitude and frequency 
inputs of the next modulator.  A guiding principle is to note that 
if we are employing signals whose instantaneous frequency is varying significantly over time, 
then we will need to account for this in the integration. From another perspective, we can also observe 
that through amplitude modulation, we are able to suppress the DC signals responsible for any carrier drift. 

We should note that, as indicated by Eq.~\ref{spectrumeq}, the resulting spectrum may be very
complex and difficult to predict in higher-order topologies. Such arrangements generally call for very
limited modulation indices. As in PM, extensions to beyond fourth-order may be of little practical 
interest.

\subsection{Operators}

In order to facilitate the design of synthesis instruments using higher-order FM topologies, we
can take advantage of the concept of an \emph{operator}. At its simplest, this is a sinusoidal
oscillator whose frequency can be modulated by another. The principle of an operator is very 
common in PM synthesis \citep{FMTheory}, and it may also include an envelope to 
allow for dynamic spectra as well as amplitude shaping. In PM, an operator is characterised
by a phase modulation input, plus amplitude and frequency parameters, and a single output. Operators 
can be connected in series (stacked), or in parallel. For FM, we can develop a similar black-box 
approach.

In order to design an operator for FM, we need to take account of our analysis in 
Section \ref{sec:theory}. We may note that within a stack, the top oscillator takes in
a modulation frequency and a modulation index, producing a modulation signal. Subsequent
oscillators take in a modulation signal in addition to the frequency and index. At the bottom of 
the stack, an oscillator produces the output signal, and it takes an amplitude instead of a modulation 
index. As in PM, the operator takes three inputs (index/amplitude and frequency scalars 
plus modulation signal). The specification requires that we make no distinction between amplitude
and index. For this to be practical, unlike in the PM case, we would then need to distinguish between 
audio and modulation outputs. For this reason, a freely-stackable FM operator requires 
two separate outputs. Envelopes may be added to shape the scalar input parameters. In pseudocode, 
the simplest design would be

\begin{lstlisting}
audio,modulation = operator(amplitude,frequency,modulation)
\end{lstlisting}

With this, the second-order stack discussed earlier would be defined as

\begin{lstlisting}
audio,mod0 = operator(index0,fm0,0)
audio,mod1 = operator(index1,fm1,mod0)
audio,mod = operator(amp,fc,mod1)
\end{lstlisting}

\begin{figure}[htp]
\begin{center}
\includegraphics[width=.5\columnwidth]{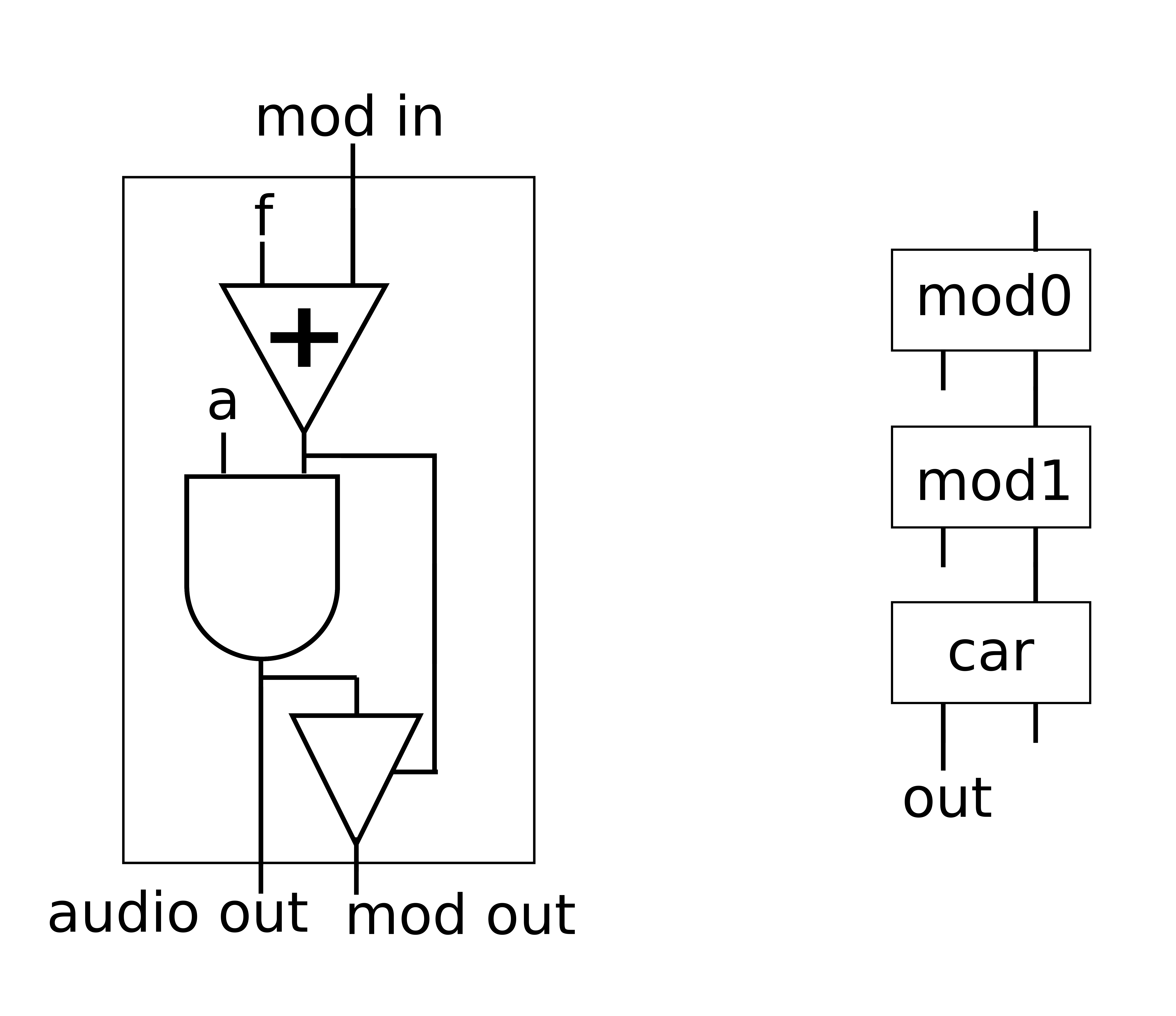}
\caption{FM operator (left) and second-order modulation arrangement (right). The \emph{a} and \emph{f}
parameters represent the scalar index/amplitude and frequency.}
\label{fig:operator}
\end{center}
\end{figure}

The implementation of the operator black box is shown in Fig~\ref{fig:operator}, together with
an arrangement of three operators in a second-order modulation topology equivalent to that
of Fig~\ref{fig:stacked}. 

As can be seen, the actual signal flow is re-ordered somewhat with the
product being placed at the output of the oscillator. This way it is possible to use a single operator
be either a carrier or a modulator in any arrangement of any order. The topmost modulator will
always have no signal inputs and we only use the audio signal out of the carrier operator. Also
we should note that this allows us to tap anywhere into an FM topology to retrieve an audio
signal at that point. Dynamic spectra can be implemented by including envelopes to control
the \emph{a} and \emph{f} parameters in Fig~\ref{fig:stacked}.

\subsection{Feedback}

\begin{figure}[htp]
\begin{center}
\includegraphics[width=.5\columnwidth]{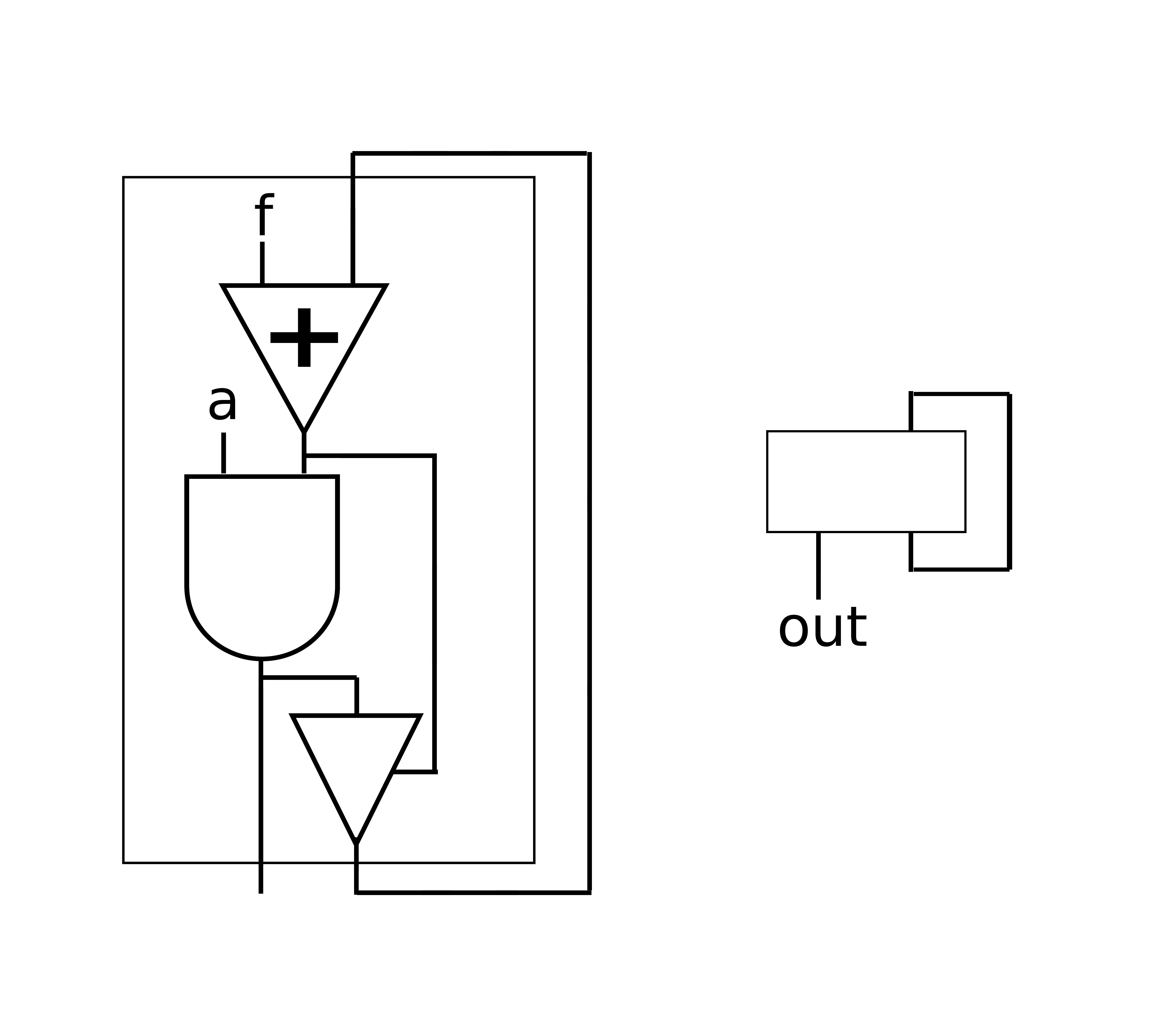}
\caption{FM operator with feedback (left) and its black-box representation (right).}
\label{fig:feedback}
\end{center}
\end{figure}

The operator as developed here opens up the possibility of implementing a feedback FM design, which 
is analogous to feedback PM as introduced by \cite{Tomisawa}. In this case, we just need to apply the 
modulation recursively to an operator, as in

\begin{lstlisting}
audio,modulation = operator(amplitude,frequency,modulation)
\end{lstlisting}

\noindent which is depicted as a flowchart in Fig~\ref{fig:feedback}.

\begin{figure}[htp]
\begin{center}
\includegraphics[width=.75\columnwidth]{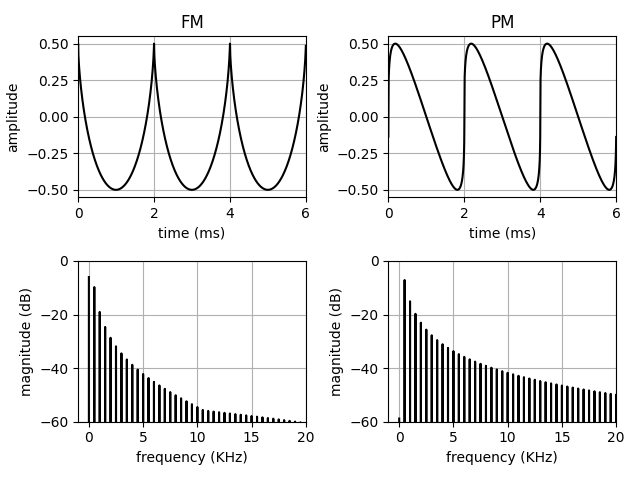}
\caption{Feedback FM (left) and PM (right) waveforms and spectra, with $f= 500$ Hz.}
\label{fig:feedbackspec}
\end{center}
\end{figure}

The spectrum and waveform of feedback FM is shown on Fig.~\ref{fig:feedbackspec} alongside
feedback PM. As we can see the two spectra share many similarities, although the waveforms are
different. This is due to the phases of each sideband, which will not be the same for the reasons
discussed earlier in Section \ref{sec:fmvpm}. 

The case of feedback PM is well understood, and the spectrum has been shown
to approximate a $1/f$ slope \citep{Lazzarini2021}.  In the case of feedback FM, there is 
a strong DC term, which we would expect as the first sideband has a cosine phase. 
The rest of the spectrum also decays with frequency following a similar curve. 
Indeed, the two signals sound very alike to the listener.

Dynamic spectra in this arrangement is possible through applying an envelope to the amplitude of
the operator. This would cause a change in both amplitude and bandwidth, as the feedback 
amount would increase at the same time as the operator output. If we want to decouple these,
it is possible to apply a separate gain to control the feedback amount, therefore varying the
number of sidebands independently from the output signal level.

Feedback FM (as opposed to Feedback PM) had been deemed not viable for implementation in
the literature \citep{PinkstonFM}. It is indeed problematic if a na\"{i}ve approach is taken, since
the integration is brought to a halt if the frequency becomes zero at any point. However, it has recently 
received a theoretical treatment by \cite{Loopback}, which shares similar principles to what we have 
developed in this article. With the proposed operator-based implementation, we have shown that it is indeed
possible to deploy the technique in practical applications. 

\subsection{Reference Implementation in C++}

The following code provides a reference implementation of the FM synthesis
operator.

\begin{lstlisting}
#include <vector>
#include <cmath>

template<typename S>
class Op {
  static constexpr long maxlen = 0x100000000; 
  const std::vector<S> &tab;  
  std::vector<S> out;
  std::vector<S> mod; 
  int lobits; 
  int lomask; 
  S lofac; 
  S fs;  
  S fac; 
  unsigned int phs;
  
  S process(S amp, int si) {
   S frac = (phs & lomask)*lofac; 
   unsigned int ndx = phs >> lobits; 
   S s = amp*(tab[ndx] + frac*(tab[ndx+1] 
   			- tab[ndx])); 
   phs += si; 
   return s;
  }

  public:
  Op(const std::vector<S> &table, float sr, 
       std::size_t vsize) :
    tab(table), out(vsize), mod(vsize), lobits(0), fs(sr), 
    fac(maxlen/sr), phs(0) {
    for(unsigned long t = tab.size()-1; 
      (t & maxlen) == 0; t <<= 1)
      lobits += 1;
    lomask = (1 << lobits) - 1;
    lofac = 1.f/(lomask + 1);
  }

  unsigned int vsize() { return out.size(); }
  S sr() { return fs; }
  
  const std::vector<S> &operator()() { return mod; }

  const std::vector<S> &operator() (S amp, S fr) {
    int si = (unsigned int) (fr*fac);
    std::size_t n = 0;
    for(auto &s : out) {
      s = process(amp, si);
      mod[n++] = s*fr;
    }
    return out;
  }

  const std::vector<S> &operator() (S amp, S fr,
  			const std::vector<S> &fm) {
    std::size_t n = 0;
    S f;
    for(auto &s : out) {
      f = fr + fm[n];
      s = process(amp, (int) (f*fac));
      mod[n++] = s*f;
    }
    return out;
  }
};
\end{lstlisting}

Using instances of this class, we can implement various types of high-order 
FM synthesis topologies. For example, a second-order stacked FM
arrangement, such as the one described in Fig.~\ref{fig:stacked}, can be 
modelled using the following class,

\begin{lstlisting}
const std::size_t def_vsize = 64;

template<typename S> class StackedFM {
  Op<S> mod0; 
  Op<S> mod1;
  Op<S> car;
  
public:
  StackedFM(const std::vector<S> &table, S fs,
            std::size_t vsize = def_vsize) :
    mod0(table,fs,vsize),mod1(table,fs,vsize),
    car(table,fs,vsize) { };

  unsigned int vsize() { return car.vsize(); }

  S fs() { return car.sr();}

  const std::vector<S> &operator()(S a,S fc,S fm0,S fm1,
				   S z0,S z1) {
    mod0(z0,fm0);
    mod1(z1,fm1,mod0());
    return car(a,fc,mod1());
  }
};
\end{lstlisting}

An object of this class can then be used to produce
an FM tone as shown by an example program fragment. 
It uses the modulation frequency ratio $c:m_0:m_1$ with separate
indices of modulation for first and second-order stages,

\begin{lstlisting}
std::vector<float> tab(1025);
std::size_t n = 0;
for(auto &s : tab)
 s = std::cos(2*M_PI/(tab.size()-1)*n++);
StackedFM<float> fm(tab,fs);

for(n = 0; n < fm.fs()*dur; n += fm.vsize()) {
 auto out = fm(amp,fr,fr,fr,z0,z1);
 for(auto s : out)
   std::cout << s << std::endl;
}
\end{lstlisting}

\section{Discussion}

While it is beyond the scope of this article to consider in detail possible applications of higher-order FM synthesis, we may cite a few. Generally, the technique is useful in situations where it is not convenient or practical to modulate the phase of a signal. One such example is the case of the synthesis methods employed in the Summit synthesiser oscillator implementation \citep{Summit}. These of course have not been published and it is not possible to exactly determine their details, but it is a reasonable assumption that PM 
is either not possible or not ideal since they opted to implement FM in a somewhat na\"{i}ve form (as per 
our earlier analysis). Another application may be found in analogue signal processing, where it is often the case that (linear) FM can be applied more directly than PM. Thus a simplification in circuit design may be also another factor 
that would favour the use of the technique.

Within a digital signal processing environment, however, there are a few practical issues arising within
the scenarios of stacked and feedback FM introduced here. These have to do with numerical errors arising 
from the discrete nature of the integration filter employed in the implementation of an oscillator. In the case of a modulation stack, we have observed that such errors result in a phase modulation term that is not present in the
continuous-time analysis of Section \ref{sec:theory}. This can introduce periodic modulation artefacts in the carrier signal that may be objectionable. In the case of Feedback FM, the errors introduce an extraneous DC term in the signal that has an obvious, although small, effect on pitch. A full analysis of these errors, their effects, and possible mitigation is beyond the scope of the present article, but it is important to note that they exist. In some applications, such as the extremely high sampling rate oscillator implementations (e.g. using field programmable gate array hardware, as in the case of Summit synthesiser, running in the MHz range), these issues may 
not be of concern.

\section{Conclusions}

A simplistic approach to implementing second and higher-order FM arrangements has been shown to have limitations. Typical issues found in these situations are related to carrier drift, which is due to the presence of a DC component in a modulating waveform. Since the amount of energy at 0 Hz is defined by the index of modulation and is a function of the Bessel coefficient associated with the relevant sideband, any timbral changes in this case are accompanied by frequency glides that may be objectionable in practical applications. These issues are fully solved through the development of a PM-equivalent form of higher-order FM. Since higher-order PM is well understood and has been successfully applied in a variety of contexts, we propose that this may be a more suitable approach.

In this paper, we defined in good detail the differences between PM and FM, demonstrating that for higher-order modulation, there needs to be some care to ensure that the correct formulation is employed. We then proposed a  econd-order FM arrangement that follows this form. From the PM-equivalent formula we are then able to describe the resulting FM spectrum by employing a similar approach to the derivation of the complex PM spectrum. 

Higher-order modulation topologies can then be implemented as an extension of the second-order approach. 
For these, we found that an operator approach may be helpful. We have then put forward the basic design of such a black box, demonstrating its equivalence to the second-order design shown earlier. With these, it is possible to freely construct various FM topologies, including feedback FM, as it is customarily done with FM. We completed the discussion with a full reference implementation of operator-based FM.

Code examples and scripts used for signal analysis in this paper can be found at\\

\begin{center}
\url{https://github.com/vlazzarini/highorderfm}
\end{center}

\bibliography{paper}
\bibliographystyle{apacite}

\end{document}